# Variance Based Scheduling Algorithm with Relay Selection And Resource Allocation in Cooperative OFDMA Networks

Maryam Basly[1], Hedia Kochkar[2] and Ammar Bouallegue[1]

[1] Université de Tunis El Manar, Ecole Nationale
d'Ingénieurs de Tunis, Sys'Com, BP 37
LeBelvedere 1002 Tunis
maryam_basly@yahoo.fr

[2] Dep. Math Informatics, National Institute of Applied
Science and Technology CentreUrbain
Nord BP 676-1080 Tunis,
hedia_kochkar@yahoo.fr

**Abstract**

Nowadays the radio interface of several standards is enhanced with advanced technologies such as OFDMA and extension technology such as relay. By using those promising transmission technology for the next generation wireless communications, scheduling problem becomes more crucial and challenging. In our work, we aim to maximize the overall system capacity while selecting the most suitable relay station under fairness constraint among both users and relay station by proposing a Gap- based scheduling. This one considers the channel state information and the unbalanced rate capacity of the two hops links. Simulations results show the effectiveness of our approach in terms of fairness and the overall system performance.
.

## 1. Introduction

Multi-hop wireless access networks are a promising solution for the high data rate coverage required in multiuser broadband wireless communications. In the last few years we have witnessed an evolution in this area from an interesting scientific research to one having a significant impact in the practice wireless communications. Several types of multi-hop wireless networks exist, such as sensor networks, ad hoc networks, and wireless mesh networks. Each of these wireless networks has different characteristics which result in different system conception like routing protocols and medium access control mechanisms.

The focus of this paper is a Relay-based system which is associated with specific base stations. The relays are used in order to extend the coverage area of the base station or increase the capacity of a wireless access system [1, 2].

Generally, it is considered that they can be used in the early phases of network deployment to ensure large coverage at a lower cost than the base station [3]. The principal task of relay is to divide one long path into several shorter one by offering alternative paths to users which are located in the shadow areas [4, 5]. In [7], the Round-Robin scheduling algorithm is extended as a performance benchmark. Two adaptive resource scheduling algorithms are proposed: greedy polling algorithm and partial proportional fair one. In [8], a proportionally fair scheduling algorithm was proposed for CDMA data network with relays. The proposed algorithm considers the instantaneous data rate and average user throughput, by considering the amount of intra-cell interference. However those researches does not deal with fairness issue and are based on scheduling scheme considered as a performance benchmark and presents no optimal solution concerning both spectral efficiency and fairness among users.

The orthogonal frequency-division multiple access (OFDMA), is considered the most promising transmission technology for the next-generation wireless communications. The resource allocation in multi-hop OFDMA system is considers as a problem challenging. There have been many studies in this area, however, only a few ones deal the resource allocation in multi-hop OFDMA systems. In [6] Kim.et al address the opportunistic sub-channel scheduling problem in relay-based OFDMA cellular networks. In this paper the BS can transmit data in both first hop link and second hop link. The opportunistic scheduling is performed not only at the BS but also at each of RSs in order to fully exploited the time varying channel state of each link, therefore the system achieve more efficient radio resource management.

In [9] a two-hop proportional fairness scheduling scheme for relay based orthogonal frequency division multiplexing access (OFDMA) systems is proposed. The proportional fairness scheduling algorithm is extended to a two-hop scenario to ensure that both direct link users and relay link users are fairly allocated capacity. Yet the proportional fair scheduling PFS is based on a priority function which is the ratio of the instantaneous data rate through the average throughput. In other words PFS priority function is based on the achieved data rate regardless of future transmission quality. [12] this paper deals with both relay selection and link scheduling. In order to maximize the network throughput in relay-assisted cellular networks, the problem is decoupled into to sub problems, a frame segmentation problem and a relay selection problem. Yet the fairness problem is not addressed in this paper, and is considered as a crucial problem especially in the downlink point to multi-points case [13]. This problem becomes even more challenging when we consider the fairness among user in both hops links since users are bandwidth hungry, the fairness issue for the scheduling algorithm should be considered at both base station (fairness among user) and relay station (fairness among relay stations).

In our paper, we consider a sub-channel allocation problem for the downlink OFDMA cellular network with fixed relay stations allowing an efficient spectral management and fairness at both BS and RS, The major contribution of the proposed scheduling algorithm lies in innovation and simplicity of the scheduler design through a adjusting decision function. This function is mainly based on the dispersion of transmission quality per set of unassigned sub channel and allows user to access a bandwidth portion taking into account the transmission quality of the next user.

The rest of this paper is organized as follows: In section 2 we describe the system model and formulates the scheduling problem. Our proposed scheduling algorithm will be presented and detailed in section 3. In section 4, we present the simulation environment, and we discuss some numerical results. In section 5 we will give some perspectives.

## 2. System Model and Problem Formulation

Let us consider an OFDMA wireless system that consists of one BS, $N$ RSs, and $K$ MSs. In this work we focus on the downlink scheduling. The total bandwidth $W$ of the system is assumed to be divided into $M$ orthogonal sub channels. Our proposed scheduling is performed through the sub channel allocation with the assumption of fixed transmission power allocation to each sub channel, previous studies have shown that this assumption may yield a marginal performance difference [11]. Since we focus on the scheduling for downlink, we denoted by T the length of a downlink sub-frame which consists of $S$ OFDMA symbols. The downlink sub-frame is divided into two slots: the first slot is used to transmit data from BS to both RSs and MSs, only in the second slot RSs can transmit data to MSs, therefore BS can transmit data in both slots. We distinguish two operations modes at the relay: The first one is considered when the scheduling at the relay station is not allowed and that data received at the RS from the base station shall immediately be forwarded to their destination. The second one is considered when the scheduling at the relay station is allowed and can be performed as in the BS or with a particular manner to RS.

The problem addressed in this paper is how assigning the available sub channel in order to meet both spectral efficiency and high aggregate throughput under fairness constraint. When the transmission is performed from base station to MSs the issue of fairness is among users. However in such system (with RSs) the issue of fairness is considered in both first hop and second hop. In the first slot, we opt to achieve fairness among direct link users and in second slot fairness among relay link users.

The channel gain of link $l$ on sub channel $n$ at sub-slot $\tau$ during scheduling slot $t$ is denoted as $h(l,n,\tau)$. We noted that the transmission power is assumed to be allocated uniformly. Of an overall manner, we can express the capacity of link $l$ on sub channel $n$ at sub-slot $\tau$ in slot $t$ as:

$$C_t(l, n, \tau) = \frac{W}{N} \log_2 \left(1 + \frac{p_n |h_t(l, n, \tau)|^2}{\Gamma\, n_0 \frac{W}{N}}\right) \quad (1)$$

Where $p_n$ is the transmission power on sub-channel n:

$$p_n = \frac{p_{tot}}{N}$$

$\Gamma = -\ln(5 \cdot BER)/1.6$ is a constant signal-to noise ratio (SNR) gap related to a target BER. $n_0 \frac{W}{N}$ is the power of additive white Gaussian noise equals to $\sigma^2$ where $n_0$ is the noise power spectral density.

From figure1, we distinguished three kinds of link, when the communication mode is direct link from BS to MSs the link capacity in this case can be expressed as:

$$C_{n,m}(t) = \frac{W}{N} \log_2 \left(1 + \frac{p_n h_t(n, m)}{\Gamma\, \sigma^2}\right) \quad (2)$$

In the event of relay communication mode the link capacity in the first sub slot (from BS to RS) can be expressed as:

$$C_{n,k}(t) = \frac{W}{N} \log_2 \left(1 + \frac{p_n h_t(n, k)}{\Gamma\, \sigma^2}\right) \quad (3)$$

In the second sub slot the communication is from RS to MS and the link capacity is

$$C_{n,k,m}(t) = \frac{W}{N} \log_2 \left(1 + \frac{p'_n h_t(n, k, m)}{\Gamma\, \sigma^2}\right) \quad (4)$$

The BS has the channel state information for both communication mode (direct link and relay link). If the communication is relay link, the BS must assign one RS for each relay link user. Let denote the set of direct link user as $D$ and the set of relay link users as $R$, the set of user served by the $RS_k$ is $R_k$ where
$k \in \{1..K\}$. Hence for a direct link user, the achieved data rate $D_m(t)$ at slot t and in the first sub slot can be expressed as:

$$D_m(t) = \sum_n^N C_{n,m} \, \alpha_{n,m} \qquad (5)$$

Where $\alpha_{n,m}=1$, if the sub channel $n$ is allocated to mobile station m, in the other case $\alpha_{n,m}=0$ where $n \in \{1..N\}$ and $m \in \{1..M\}$.
The data rate of the $RS_k$ in the first sub slot is:

$$D_{RS,\tau=1}(k) = \sum_{n=1}^N C_{n,k} \, \alpha_{n,k} \qquad (6)$$

if the sub channel $n$ is allocated to the $k^{th}$ relay station then $\alpha_{n,k}=1$ else $\alpha_{n,k}=0$, where, $n \in \{1..N\}$ and $k \in \{1..K\}$.
In the second sub slot the achieved data rate is expressed as follows:

$$D_{RS,\tau=2}(k) = \sum_{m=1}^{M'} C_{n,k} \, D_{m,k,\tau=2} \qquad (7)$$

Where $M'$ is the cardinality of $R_k$. and $D_{m,k}$ is the achieved data rate of relay link user $m$ at $RS_k$ and in the second sub slot.

$$D_{m,k,\tau=2} = \sum_n^N C_{n,k,m} \, \beta_{n,k,m} \qquad (8)$$

Where $\beta$ for the second hop slot link is used to indicate whether sub channel $n$ is allocate to user $m$ via $RS_k$ .hence the achieved data rate for the relay link user $m$ in the first time sub slot is

$$D_{m,k,\tau=1} = \frac{D_{RS,\tau=1}}{D_{RS,\tau=2}} * D_{m,k} \qquad (9)$$

Due to the asymmetry of two links hops the effective data rate of one relay link user can be expressed as

$$D_{eff,m} = \min\{D_{m,k,\tau=1}, D_{m,k,\tau=2}\}, \text{ where } m \in R \qquad (10)$$

Then the system capacity can be expressed as follow:

$$C_{Sys}(t) = C_{SysDL}(t) + C_{SysRL}(t) \qquad (11)$$

The first term in this equation is for the direct link and the last one is for relay link.

$$C_{SysDL}(t) = \sum_{n=1}^N \sum_{m=1}^M \sum_{\tau=1}^2 C_{n,m} \alpha_{n,m}(t) \qquad (12)$$

$$C_{SysRL}(t) = \sum_{k=1}^K \sum_{m=1}^{M'} D_{eff}(t) \qquad (13)$$

Having the target to maximize the system capacity, the objective function during one frame can be formulated as follows:

$$\forall T, C_{Sys} = \sum_{t=1}^T \left( \sum_{n=1}^N \sum_{m=1}^M \sum_{\tau=1}^2 C_{n,m}(t)\alpha_{n,m}(t) + \sum_{k=1}^K \sum_{m=1}^{M'} D_{eff}(t) \right) \qquad (14)$$

Subject to:

$$\begin{cases} C_1 & \alpha_{n,m} \in \{0,1\} \; \forall n \in N \quad \forall m \in D \\ C_2 & \sum_{n=1}^N \alpha_{n,m} = 1, \quad \forall n \in N \\ C_3 & \sum_{\tau=1}^2 \sum_{n=1}^N \alpha_{n,m} \leq 1, \quad \forall m \in D \\ C_4 & \overline{R_1} \approx \overline{R_2} .... \approx \overline{R_m} \quad \forall m \in D \cup R \end{cases}$$

The constraints $C_1$ and $C_2$ have the objective to ensure that each sub channel is assigned to only one *MS* where *N* and *D* denote, respectively, the set of sub channels in the cell and the set of direct link user. The constraint $C_3$ denotes that one *MS* could have only one sub channel at the same time. The last one $C_4$ corresponds to the fairness constraint and is among direct link users and relay link users, where $R_m$ is the achieved data rate for mobile station *m* and can be update according the used communication mode as in equation *(5)* and *(10);* so the average data rate will be updated as following:

$$\overline{R_m}(t) = \left(1 - \frac{1}{T}\right) \times \overline{R_m}(t-1) + D_m(t)/T \qquad (15)$$

## 3. Proposed scheduling algorithm

In opportunistic MaxSnr scheduling algorithm, the time-varying properties of wireless channels are exploited for efficient use of radio resources. Since mobile stations have different channel states from each other in each time slot, MaxSnr scheduling scheme allocates more resources to MSs that presents a better channel state at each transmission time. By doing, so the spectral efficiency could be improved while achieving multiuser diversity in wireless. Nevertheless, this approach does not worry about the transmission quality of the next user. To explain the weakness of this approach and the main idea of the proposed scheduling algorithm, we consider a simplified scenario, where two users share two sub channels as presented in the following:

Table 1: Simplified Scenario

|  | Sub-channel 1 | Sub-channel 2 |
|---|---|---|
| User 1 | 60 | 10 |
| User 2 | 80 | 70 |

Let suppose that the *SNR (signal to noise ratio)* reflect the transmission quality of each sub channel. According Table 1 for example the first sub channel presents a better transmission rate for user 1than user 2.

If MaxSnr is used, user with the best transmissible rate is allowed to transmit first then user 2 will benefit the sub-channel 1 and the user 1 will use sub-channel 2, the system receive 80+10=90, by doing so fairness among user is not taken into account. The main idea of using variance is to allowed user with high variance to transmit first. The variance metric is expressed as:

$$V(k) = \frac{\sum_{i=1}^{M} \delta_{k,i}^2}{M} - \left(\frac{\sum_{i=1}^{M} \delta_{k,i}}{M}\right)^2 \qquad (16)$$

Where δ is the signal to noise ratio of each sub-channel. Based on the assumption of power control and Rayleigh fading, δ is modeled by an i.i.d exponential distribution.
The variance of user 1 and user 2 according the previous example is respectively 1250 and 50. Since user1 presents a high variance than user 2, it will be allowed to transmit firstly. With this manner the system receives 60+70=130, we may find an optimal solution for both spectral efficiency and fairness. The priority function can be expressed as:

$$Pm = Vm \qquad (17)$$

Where $V_m$ is the *SNR* variance of user *m* according unassigned sub channel. User who satisfies the following equation is allowed to send before the other.

$$m^* = \arg\max \{V1, V2, ...VM\} \qquad (18)$$

At this stage we have just selected the adequate user for both spectral efficiency and fairness among users; in the following we will explain how to select the adequate relay station in order to meet fairness among the RSs firstly then how to choose the communication mode in order to maximize the overall system throughput.

The selected user $m^*$ has two options, direct link communication mode or relay link communication mode. If the communication mode is a direct link the $m^*$ is allowed to use its best sub channel that verified the equation bellow:

$$n^* = \arg\max \{\delta_1, \delta_2, ...\delta_N\} \qquad (19)$$

Where $\delta_i$ is the *SNR* of the $i^{th}$ sub-channel from *BS* to *MS* as shown in figure 1.

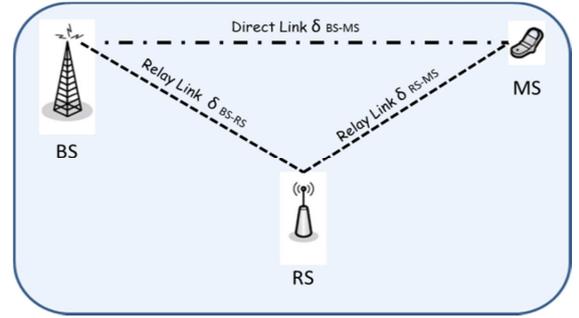

Figure 1      Direct and Relay communication mode

The next step of the proposed scheduling algorithm is to select the adequate RS and is done by the set of RSs which would report to the BS the *SNR* of each sub-channel. In this work the effect of the distance between MS and RS is neglected, we assume that the *SNR* reflect perfectly the transmission quality. Then the BS selects RS according their variance metric, similarly to the users' process selection, the RS with high variance is allowed to be used at first in order to meet fairness among relay station. Once the relay station is chosen, the second option is when the communication mode is a relay mode. According the equation (10) the transmission rate of MS $m^*$ can be expressed as min {$\delta_{BS-RS}$, $\delta_{RS-BS}$}. In order to maximize the overall system throughput under fairness constraint, we are proposing to choose the communication mode which meets the following statement: argmax{ $\delta_{BS-MS}$, min {$\delta_{BS-RS}$, $\delta_{RS-BS}$}}.

## 4. Numerical Results And Performance Evaluation

To illustrate the advantage of our proposed Relay-variance scheduling algorithm, we compared it to Relay-MaxSnr. The system considered for simulation is the downlink of an OFDMA based relay enhanced cellular network with one centered BS, 6 uniformly distributed RSs and 10 MSs. The used bandwidth is 10 Mhz partitioned into 128 sub-channels in each hop link. Figure2 shows the overall system throughput for both Relay-MaxSnr and Relay-Variance algorithm versus simulation time. The Relay-MaxSnr promotes user with the best sub-channel transmission rate in both hops links, whereas in Relay-Variance algorithm, user with a significant dispersion in its channel transmission quality is allowed to transmit firstly, with this manner and as explain in the previous scenario we can achieve a better spectrum management which induces in an improvement in the system throughput.

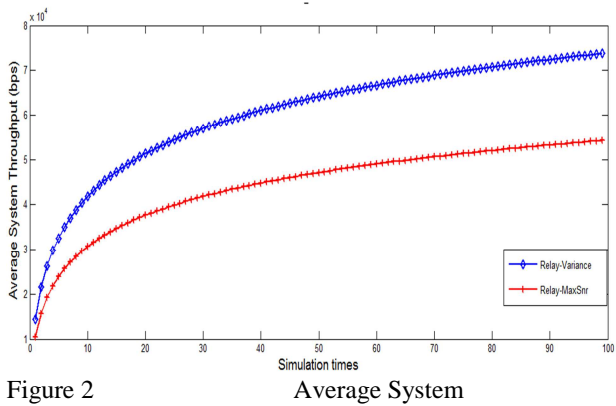

Figure 2            Average System

Another benefit of using the variance related to the SNR of the set of concerning sub-channels is the manner to select the relay station when the communication mode is not a direct link, indeed when we select the relay station that present the maximum variance during the second hop we can achieve both good transmission rate for this relay station and fairness among the set of relay. In figure 3, in order to show the fairness degree among RSs we simulate the Jain's fairness index according the RSs set.

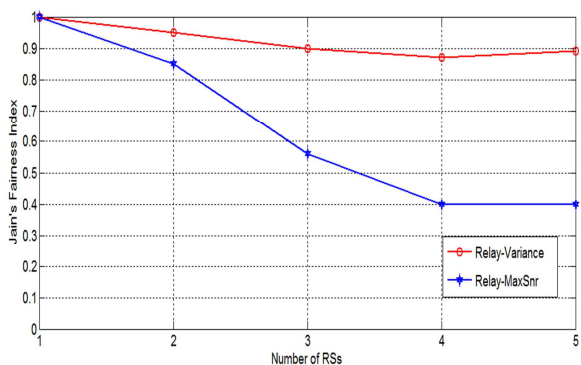

Figure 3            Jain's Fairness Index Versus RSs

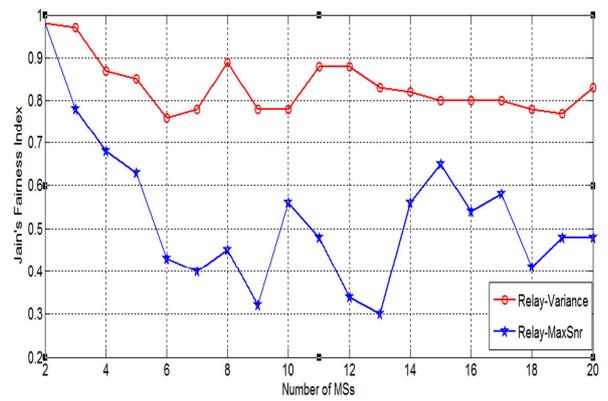

Figure 4            Jain's Fairness Index Versus MSs

It is clearer that the jain's index remains near the optimal case with 0.97 for Relay-variance algorithm but we observe a significant degradation for the Relay-MaxSnr scheme. In figure 4 the Jain's fairness index is among users. As users with significant variance have a high probability to missing their best sub channels, they have high priority to transmit first than the other. This manner can improve significantly the achieved data rate while maintaining the fairness among users.

## 5. Conclusions

In this paper, we proposed a novel fair variance resource allocation scheme for relay OFDMA cellular networks. To solve the resource allocation problem under fairness constraints, we propose a priority function based variance process. This process is used to select both user and relay station. Numerical results have shown that our proposed scheme can improve significantly the overall system performance and the fairness among users and relay stations. Further work have be done in order to improve our scheme such as considering different class of traffic and taking into account QoS requirement in terms of packet loss rate and delay.